\documentclass[12pt,preprint,notoc,nohyper]{MCCYL} 

\usepackage{epsfig,multicol,bbm}

\usepackage{rotating}

\newcommand\fverb{\setbox\pippobox=\hbox\bgroup\verb}
\newcommand\fverbdo{\egroup\medskip\noindent%
            \fbox{\unhbox\pippobox}\ }
\newcommand\fverbit{\egroup\item[\fbox{\unhbox\pippobox}]}
\newbox\pippobox

\title{Do the Ricci and energy-momentum tensors have
``duality'' in the context of their Lie symmetries?}

\author{Hina Khan$^1$, Asghar Qadir$^2$, K. Saifullah$^3$ and M.
Ziad$^4$ \\

$^1$Department of Sciences and Humanities,
National University of Computer and Emerging Sciences, Islamabad, Pakistan \\

$^2$Centre for Advanced Mathematics and Physics,
National University of Sciences and Technology, Rawalpindi, Pakistan\\

$^3$Department of Mathematics, Quaid-i-Azam University,
Islamabad, Pakistan\\

$^4$Department of Mathematics and Statistics, Sultan Qaboos
University, Oman \\

Electronic address: \email{khan\_hina22@hotmail.com},
\email{aqadirmath@yahoo.com}, \email{saifullah@qau.edu.pk},
\email{mziad@squ.edu.om}}

\preprint{}  

\abstract{The Ricci and energy-momentum tensors have the same
algebraic symmetries. In the Einstein equations they look ``dual''
to each other, in that interchanging them and inverting the
gravitational coupling leaves the equations invariant. It may then
be expected that their differential symmetry Lie algebras would also
be identical. Using cylindrically symmetric static spacetimes it is
shown that they are not identical and neither algebra is a subset of
the other.}

\begin{document}

\section{Introduction}

Lie symmetries of various geometrical and physical quantities in
general relativity have been studied for some time \cite{exactsoln,
hallbook}. Isometries, or Killing vectors (KVs), the vector fields
along which the metric tensor, $\mathbf{g}$, remains invariant under
Lie transport, have been used to construct new solutions of the
Einstein field equations (EFEs)
\begin{equation}
R_{ab}-\frac{1}{2}Rg_{ab}=\kappa T_{ab};    (a,b=0,1,2,3)
\label{EFE}
\end{equation}
where $\mathbf{R}$ is the Ricci tensor, $\mathbf{T}$ the
energy-momentum tensor, $R$ the Ricci scalar and $\kappa$ the
gravitational coupling $8\pi G/c^{4}$. While the KVs give the
symmetries inherent in the space itself, invariance under the Lie
transport of the energy-momentum tensor gives the symmetries of the
matter content of the space (called matter collineations or MCs
\cite{carot,hall}), and hence is more relevant physically. Since it
appears in the EFEs with the Ricci tensor, the symmetries of the
Ricci tensor (called Ricci collineations or RCs \cite{exactsoln,
hallbook}) are also physically relevant. These vector fields also
provide invariant bases for the classification of the solutions of
the EFEs. A vector field $\mathbf{\xi }$ is an MC if the Lie
derivative of the energy-momentum tensor vanishes along
$\mathbf{\xi}$

\begin{equation}
{\Large \pounds }_{\mathbf{\xi }}\mathbf{T}=0.  \label{LR}
\end{equation}
In component form, the MC equation (\ref{LR}) takes the form

\begin{equation}
\xi ^{c}T_{ab,c}+T_{ac}\xi _{,b}^{c}+T_{bc}\xi _{,a}^{c}=0.
\label{RC}
\end{equation}
Here comma denotes a partial derivative with respect to the
coordinates. For four dimensional space these are ten coupled
partial differential equations which are to be solved for the four
components of the vector $\mathbf{\xi =} (\xi ^{0},\xi ^{1},\xi
^{2},\xi ^{3})$. If the energy-momentum tensor in the last equation
is replaced by the Ricci tensor, the vector field is an RC, and if
it is replaced by the Riemann curvature tensor, the vector field is
called a curvature collineation (CC) \cite{exactsoln, hallbook}. It
is well known that every KV is a CC and every CC, in turn, is an RC
but the converse is not true in general. Mutual relationships
between different spacetime symmetries are represented graphically
in the inclusion diagram in Ref. \cite{katzin}. An interesting
question arises here about the place of MCs in this diagram, and
that is the subject of this paper.

There has been recent interest in the study of RCs of plane
symmetric \cite{15}, spherically symmetric  \cite{sp1},
cylindrically symmetric \cite{KS} and various other classes of
spacetimes \cite{RC}. As mentioned earlier, if the components of the
energy-momentum tensor, $T_{ab}$, in Eq.(\ref{RC}) are replaced by
those of the Ricci tensor, we get RCs. Due to the similarity of the
mathematical form of the Ricci and the energy-momentum tensors, and
similarity of their collineation equations, attempts were made to
obtain the results for MCs from those of RCs \cite{MC, QS} by
replacing the Ricci tensor by the energy-momentum tensor. These
attempts assume that because of the identity of the algebraic
symmetries of the two tensors their differential symmetries would
also be identical i.e. their corresponding algebras would be same.
In this paper we show that this is not true. Another aim of this
paper is to investigate the next and more important question of
``duality'' between these two tensors, as regards their
collineations. To achieve this, we construct all possible
(inclusion) relationships between RCs and MCs. We find that
cylindrically symmetric static spacetimes provide a very useful
framework for this investigation, as all the components of the Ricci
tensor are independent (which is not the case in spherical symmetry,
for example). This fact gives rise to a whole lot of possibilities
for the relationship between RCs and MCs. We investigate all these
possibilities here and demonstrate that there is no inclusion
relation between the two algebras.

The plan of the paper is as follows. In the next section we give the
MC equations and discuss their solution. In Section 3 all the
possibilities of relationships between MCs and RCs are identified
and specific examples provided for each of these cases. The
concluding remarks are given in Section 4. Tables 1-5 which
summarize the solutions of the MC equations are provided in the
Appendix.

\section{Matter collineations of cylindrically symmetric static spacetimes}

The line element for the general cylindrically symmetric static
spacetimes in $\left( t,\rho ,\theta ,z\right) $ coordinates can be
written as \cite{exactsoln}

\begin{equation}
ds^{2}=e^{\nu \left( \rho \right) }dt^{2}-d\rho ^{2}-a^{2}e^{\lambda
\left( \rho \right) }d\theta ^{2}-e^{\mu \left( \rho \right)
}dz^{2}, \label{1}
\end{equation}
where the minimal symmetry is given by the three Killing vectors,
$\partial _{t}$, $\partial _{\theta }$, $\partial _{z}$. For this
metric the only non-zero components of the Ricci tensor are
\begin{equation}
\left.
\begin{array}{l}
R_{00}=\frac{e^{\nu }}{4}\left( 2\nu ^{\prime \prime }+\nu ^{\prime
^{2}}+\nu ^{\prime }\lambda ^{\prime }+\nu ^{\prime }\mu ^{\prime }\right), \\
R_{11}=-\left( \frac{\nu ^{\prime \prime }}{2}+\frac{\lambda
^{\prime \prime }}{2}+\frac{\mu ^{\prime \prime }}{2}+\frac{\nu
^{\prime ^{2}}}{4}+\frac{\lambda ^{\prime ^{2}}}{4}+\frac{\mu ^{\prime ^{2}}}{4}\right), \\
R_{22}=-\frac{a^{2}e^{\lambda }}{4}\left( 2\lambda ^{\prime \prime
}+\nu ^{\prime }\lambda ^{\prime }+\lambda ^{\prime ^{2}}+\lambda
^{\prime }\mu
^{\prime }\right), \\
R_{33}=-\frac{e^{\mu }}{4}\left( 2\mu ^{\prime \prime }+\nu ^{\prime
}\mu ^{\prime }+\lambda ^{\prime }\mu ^{\prime }+\mu ^{\prime
^{2}}\right).
\end{array}
\right.  \label{Ricci}
\end{equation}
Here $^{,\prime ,}$\ denotes differentiation with respect to $\rho$.
The Ricci scalar is given by
\begin{equation}
R=\nu ^{\prime \prime }+\lambda ^{\prime \prime }+\mu ^{\prime
\prime }+ \frac{1}{2}\left( \nu ^{\prime ^{2}}+\lambda ^{\prime
^{2}}+\mu ^{\prime ^{2}}+\nu ^{\prime }\lambda ^{\prime }+\nu
^{\prime }\mu ^{\prime }+\lambda ^{\prime }\mu ^{\prime }\right) .
\end{equation}
Using the EFEs (Eq. \ref{EFE}), the general form of the
energy-momentum tensor, $T_{b}^{a}$, becomes
\begin{equation}
\left.
\begin{array}{l}
T_{0}^{0}=-\frac{1}{4}\left( 2\lambda ^{\prime \prime }+2\mu
^{\prime \prime }+\lambda ^{\prime ^{2}}+\mu ^{\prime ^{2}}+\lambda
^{\prime }\mu ^{\prime
}\right), \\
T_{1}^{1}=-\frac{1}{4}\left( \nu ^{\prime }\lambda ^{\prime }+\nu
^{\prime
}\mu ^{\prime }+\lambda ^{\prime }\mu ^{\prime }\right), \\
T_{2}^{2}=-\frac{1}{4}\left( 2\nu ^{\prime \prime }+2\mu ^{\prime
\prime
}+\nu ^{\prime ^{2}}+\mu ^{\prime ^{2}}+\nu ^{\prime }\mu ^{\prime }\right),\\
T_{3}^{3}=-\frac{1}{4}\left( 2\nu ^{\prime \prime }+2\lambda
^{\prime \prime }+\nu ^{\prime ^{2}}+\lambda ^{\prime ^{2}}+\nu
^{\prime }\lambda ^{\prime }\right).
\end{array}
\right.  \label{matter}
\end{equation}

Now, the solution of Eqs. (\ref{RC}) for the energy-momentum tensor
is similar to the one given in Ref. \cite{KS}, and it can be written
simply by replacing the components of the Ricci tensor there by
those of the energy-momentum tensor. Therefore, we will not give the
MC vectors for different cases and their corresponding Lie algebras
and Lie groups here again and the reader is referred to Ref.
\cite{KS} for all these details. However, we will reproduce the
tables of the main results here as we will need to refer to them
frequently in the next section. It may be pointed out here again
that during the course of solution of the MC (or RC) equations one
gets different cases which are characterized by the constraints on
the components of the energy-momentum (or Ricci) tensor. We will be
using the same notation and case numbering here as used in Ref.
\cite{KS} for easy comparison. In fact, if we solve Eqs. (\ref{RC})
for a general second rank, symmetric and diagonal tensor $A_{ab}$,
we not only get the KVs \cite{QZ} and RCs \cite{KS} for
cylindrically symmetric static spacetimes but also find the MCs
explicitly. This means that these tables can be used to obtain
complete information on these three symmetries. There is one point
however that while the Ricci and the energy-momentum tensors can be
degenerate (i.e. the determinant is zero) as well as non-degenerate
(i.e. the determinant is non-zero), the metric tensor cannot be
degenerate. We see that when the Ricci tensor is non-degenerate, the
Lie algebra of the RCs is always finite-dimensional. However, when
it is degenerate, it admits a finite-dimensional Lie algebra only
when $R_{11}=0$, $R_{ii}\neq 0$, $i=0,2,3$. This holds for MCs also.
Tables 1-5 are for finite-dimensional Lie algebras only. The numbers
in the last column indicate the dimension of the Lie algebra
admitted by $\mathbf{\xi }$ and equation numbers there refer to
those in Ref. \cite{KS}. Further, as we are dealing with diagonal
tensors, for simplicity we will write $R_{i}$ and $T_{i}$ for
$R_{ij}$ and $T_{ij} (i=j)$, respectively.

\section{Matter and Ricci Collineations}

In what follows we write ``finite (or infinite) MCs'', in place of
``MCs having finite (or infinite) dimensional Lie algebra'', for the
sake of brevity. Similarly, we write ``non-degenerate (or
degenerate) MCs'' when we mean ``MCs for the non-degenerate (or
degenerate) energy-momentum tensor''. The same holds for RCs also.
We find that depending upon whether the MCs and RCs are degenerate
or non-degenerate, finite or infinite, all possible relationships
between them can be written in the form of the following table,
where the last column gives the example number for the corresponding
case.

\bigskip

\begin{flushleft}

\begin{tabular}{|l|l|l|l|l|}
\hline
\multicolumn{5}{|c|}{\textbf{Possible relationships between MCs and RCs}} \\
\hline Non-Degenerate MCs & (Finite MCs) & Non-Degenerate RCs &
(Finite RCs) & 3.1 \\ \hline &  & Degenerate RCs & Finite RCs & 3.2
\\ \hline &  &  & Infinite RCs & 3.3 \\ \hline
Degenerate MCs & Finite MCs & Non-Degenerate RCs & (Finite RCs) &
3.4
\\ \hline
&  & Degenerate RCs & Finite RCs & 3.5$^*$ \\ \hline &  &  &
Infinite RCs & 3.6$^*$ \\ \hline & Infinite MCs & Non-Degenerate RCs
&
(Finite RCs) & 3.7 \\ \hline &  & Degenerate RCs & Finite RCs & 3.8 \\
\hline &  & & Infinite RCs & 3.9 \\ \hline
\multicolumn{5}{|c|}{$^*$Examples for these cases have not been provided} \\
\hline
\end{tabular}
\end{flushleft}

\bigskip

The metrics for all these possibilities have been constructed with
the exception of two cases. The examples of these spacetimes given
below also demonstrate the procedure of finding MCs, RCs and KVs
from Tables 1-5 in the Appendix. We shall call MCs (or RCs)
\emph{proper} if they are not KVs.

\subsection{Non-degenerate (finite) MCs; non-degenerate (finite) RCs}

In this case both the energy-momentum and the Ricci tensor are
non-degenerate having finite MCs and RCs. Consider the metic

\begin{equation}
ds^{2}=\cosh ^{2}k\rho dt^{2}-d\rho ^{2}-a^{2}\left( \cosh k\rho
\right) ^{-1}d\theta ^{2}-\left( \cosh k\rho \right) ^{-1}dz^{2},
\end{equation}
For this metric the components of energy-momentum tensor are
\begin{eqnarray*}
T_{0} &=&\cosh ^{2}k\rho \frac{k^{2}}{4}\left( 4-7\tanh ^{2}k\rho
\right), \\
T_{1} &=&-\frac{3k^{2}}{4}\tanh ^{2}k\rho, \\
T_{2} &=&a^{2}\left( \cosh k\rho \right) ^{-1}\frac{k^{2}}{4}\left(
2+\tanh
^{2}k\rho \right) \;, \\
T_{3} &=&\left( \cosh k\rho \right) ^{-1}\frac{k^{2}}{4}\left(
2+\tanh ^{2}k\rho \right) \;.
\end{eqnarray*}
This is an anisotropic fluid with energy density positive for $0\leq \rho <%
\frac{1}{k}\tanh ^{-1}\frac{2}{\sqrt{7}}$ and negative for $\rho \geq \frac{1%
}{k}\tanh ^{-1}\frac{2}{\sqrt{7}}$. However, with a cosmological
constant greater than $\frac{3}{4}k^{2}$, the energy density becomes
positive definite. For this metric the components for $R_{ab}$ are
\begin{eqnarray*}
R_{0} &=&k^{2},
R_{1} =-\frac{3k^{2}}{2}\tanh ^{2}k\rho , \\
R_{2} &=&\frac{k^{2}}{2}\left( \sec hk\rho \right) ^{3}, R_{3}
=\frac{k^{2}}{2}\left( \sec hk\rho \right) ^{3}.
\end{eqnarray*}
It admits 4 MCs (Case AIIa(2)), 7 RCs (Case BIVb3(ii)$\gamma _{2}$)
and 4 KVs and, therefore, is a case of proper RCs.

\subsection{Non-degenerate (finite) MCs; degenerate and finite RCs}

Here we provide an example of a metric with non-degenerate
energy-momentum tensor and degenerate Ricci tensor with both MCs and
RCs finite. Consider

\begin{equation}
ds^{2}=\left( \rho /\rho _{0}\right) ^{2a}dt^{2}-d\rho ^{2}-\left(
\rho /\rho _{0}\right) ^{2b}\alpha ^{2}d\theta ^{2}-\left( \rho
/\rho _{0}\right) ^{2c}dz^{2} ,
\end{equation}
where, $a=(1\pm \sqrt{3})/2$, $b=c=1/2$, and one gets $R_{1}=0$ and
$R_{i}$ are non-zero constants for $i=0,2,3$. For $a=(1+\sqrt{3}
)/2$, we have $T_{ab}\,$ in component form
\begin{eqnarray*}
T_{0} &=&\frac{\rho ^{\sqrt{3}-1}}{4\rho ^{\sqrt{3}+1}} ,
T_{1} =(\frac{3}{4}+\frac{\sqrt{3}}{2})\rho ^{-2} , \\
T_{2} &=&-\frac{\alpha ^{2}(2+\sqrt{3})}{4\rho \rho _{0}} ,  T_{3}
=-\frac{2+\sqrt{3}}{4\rho \rho _{0}}\;.
\end{eqnarray*}
For this metric $R_{ab}$ has the following components
\begin{eqnarray*}
R_{0} &=&\frac{(1+\sqrt{3})^{2}\rho ^{\sqrt{3}-1}}{4\rho
_{0}^{\sqrt{3}+1}},
R_{1} =0 , \\
R_{2} &=&-\frac{\,\alpha ^{2}(1+\sqrt{3})}{4\rho _{0}\rho }, R_{3}
=-\frac{\,(1+\sqrt{3})}{4\rho _{0}\rho }.
\end{eqnarray*}
It admits 5 MCs (AIIb1(i)$\beta $), 5 RCs (Case IIBd4(i)) and 4 KVs
(Case AIIa(2)), and therefore, is a case of proper MCs and RCs.

\subsection{Non-degenerate (finite) MCs; degenerate and infinite RCs}

Here we discuss the example of non-degenerate energy-momentum and
degenerate Ricci tensors with finite MCs but infinite dimensional RC
algebra.

\begin{equation}
ds^{2}=e^{A\rho }\left( dt^{2}-dz^{2}\right) -d\rho
^{2}-a^{2}d\theta ^{2},
\end{equation}
$A$ is a non-zero constant. For this metric the components of
$T_{ab}\,$are
\begin{eqnarray*}
T_{0} &=&-\frac{e^{A\rho }A^{2}}{4} ,
T_{1} =\frac{A^{2}}{4}\;, \\
T_{2} &=&\frac{3a^{2}A^{2}}{4} , T_{3} =\frac{e^{A\rho }A^{2}}{4}\;.
\end{eqnarray*}
$R_{ab}$ has the following components
\begin{eqnarray*}
R_{0} &=&\frac{e^{A\rho }A^{2}}{2} ,
R_{1} =-\frac{A^{2}}{2}\;, \\
R_{2} &=&0 ,  R_{3} =-\frac{e^{A\rho }A^{2}}{2} .
\end{eqnarray*}
It has 7 MCs (Case AIa1(i)), RCs have infinite dimensional Lie
algebra (Case (III)) and 7 KVs (Case AIa1(i)). It is anti-Einstein
and anisotropic with negative energy.

\subsection{Degenerate and finite MCs; non-degenerate (finite) RCs}

One of the examples of metrics with degenerate energy-momentum
tensor with finite MCs and non-degenerate Ricci tensor with finite
RCs is provided here.

\begin{equation}
ds^{2}=\left( \rho /\rho _{0}\right) ^{-1/2}dt^{2}-d\rho ^{2}-\left(
\rho /\rho _{0}\right) \alpha ^{2}d\theta ^{2}-\left( \rho /\rho
_{0}\right) dz^{2}
\end{equation}
Taking $a=-1/4$, $b=c=1/2$ in metric (A3) gives the above metric.
For this metric the components of $T_{ab}$ are
\begin{eqnarray*}
T_{0} &=&\frac{\rho _{0}^{1/2}}{4\rho ^{5/2}} ,
T_{1} =0 , \\
T_{2} &=&-\frac{\alpha ^{2}}{16\rho \rho _{0}} , T_{3}
=-\frac{1}{16\rho \rho _{0}} .
\end{eqnarray*}
and components of $R_{ab}$ are
\begin{eqnarray*}
R_{0} &=&\frac{\rho _{0}^{1/2}}{16\rho ^{5/2}} ,
R_{1} =\frac{3}{16\rho ^{2}}, \\
R_{2} &=&\frac{\alpha ^{2}}{8\rho \rho _{0}} , R_{3} =\frac{1}{8\rho
\rho _{0}} .
\end{eqnarray*}
It has 5 MCs (Case II Bd4(i)), 5 RCs (Case AIIb1(i)$\beta $) and
4KVs (Case AIIa(2)).

\subsection{Degenerate and finite MCs; degenerate and finite RCs}

This is the case where an example has eluded our attempts. It would
really be interesting to see if an example of a spacetime exists for
which both the RCs and MCs are finite and degenerate. The necessary
conditions for this to happen are that $R_{11}$ and $T_{11}$ are
zero and other components non-zero. Alternatively, a proof of
non-existence of such a space would also be very interesting.

\subsection{Degenerate and finite MCs; degenerate and infinite RCs}

Here also we have not been able to find an example for which both
the MCs and RCs are degenerate but the former is finite and the
latter is infinite. But this is where the question of ``duality"
between the energy-momentum and the Ricci tensors become important
because we have its ``mirror" example in Case 3.8, where although
both are degenerate the MCs are infinite while the RCs are finite.
Non-existence of an example here would imply that there is no
``duality" between the two tensors as far as their collineations is
concerned.

\subsection{Degenerate and infinite MCs; non-degenerate (finite) RCs}

Here we provide a metric with infinite dimensional MC algebra and
finite RCs.

\begin{equation}
ds^{2}=\left( \rho /\rho _{0}\right) ^{2a}dt^{2}-d\rho ^{2}-\left(
\rho /\rho _{0}\right) ^{4/3}\alpha ^{2}d\theta ^{2}-\left( \rho
/\rho _{0}\right) ^{4/3}dz^{2}
\end{equation}
Choosing $a\neq 4/3$, $0$ , $2/3$ , $-1/3$ , $1$ and $b$ $=c=2/3$ in
metric (A3) gives above metric. For this metric the components of
$T_{ab}\,$are
\begin{eqnarray*}
T_{0} &=&0 ,
T_{1} =\frac{4(a+1/3)}{3\rho ^{2}}, \\
T_{2} &=&-\frac{\alpha ^{2}(3a-9a^{2}+2)}{9\rho ^{2/3}\rho
_{0}^{4/3}} ,  T_{3} =-\frac{(3a-9a^{2}+2)}{9\rho ^{2/3}\rho
_{0}^{4/3}} .
\end{eqnarray*}
and $R_{ab}$ has the following components
\begin{eqnarray*}
R_{0} &=&\rho _{0}^{-2a}a(a+1/3)\rho ^{2a-2} ,
R_{1} =-\frac{(9a^{2}-9a-4)}{9\rho ^{2}} , \\
R_{2} &=&-\frac{2\alpha ^{2}(3a+1)}{9\rho _{0}^{4/3}\rho ^{2/3}} ,
R_{3} =-\frac{2(3a+1)}{9\rho _{0}^{4/3}\rho ^{2/3}} .
\end{eqnarray*}
It admits infinite dimensional MCs (Case (I)), 5 RCs (CaseA
IIb1(i)$\beta $) and 4 KVs (Case AIIa(2)).

\subsection{Degenerate and infinite MCs; degenerate and finite RCs}

Here both the energy-momentum and the Ricci tensors are degenerate
with finite RCs and infinite MCs.

\begin{equation}
ds^{2}=\left( \rho /\rho _{0}\right) ^{8/3}dt^{2}-d\rho ^{2}-\left(
\rho /\rho _{0}\right) ^{4/3}\alpha ^{2}d\theta ^{2}-\left( \rho
/\rho _{0}\right) ^{4/3}dz^{2}\,.
\end{equation}
Taking $a=4/3$, $b=c=2/3$ in metric (A3) gives the above metric. The
components of $T_{ab}$ are
\begin{eqnarray*}
T_{0} &=&0 ,
T_{1} =\frac{20}{9}\rho ^{-2} , \\
T_{2} &=&\frac{10\alpha ^{2}}{9\rho ^{2/3}\rho _{0}^{4/3}} , T_{3}
=\frac{10}{9\rho ^{2/3}\rho _{0}^{4/3}} .
\end{eqnarray*}
for this metric $R_{ab}$ has the following components
\begin{eqnarray*}
R_{0} &=&\frac{20\rho ^{2/3}}{9\rho _{0}^{8/3}} ,
R_{1} =0 , \\
R_{2} &=&-\frac{10\alpha ^{2}}{9\rho ^{2/3}\rho _{0}^{4/3}} , R_{3}
=-\frac{10}{9\rho ^{2/3}\rho _{0}^{4/3}} .
\end{eqnarray*}
It admits MCs having infinite dimensional Lie algebra (CaseI), 5 RCs
(Case IIBd4(i)) and 4 KVs (AIIa(2)).

\subsection{Degenerate and infinite MCs; degenerate and infinite RCs}

The case of infinite dimensional algebras for both the MCs and RCs
when the two tensors are degenerate is discussed here.

\begin{equation}
ds^{2}=cosh^{2}(A+B\rho )dt^{2}-d\rho ^{2}-a^{2}d\theta ^{2}-dz^{2},
\end{equation}
$A$ and $\alpha $ are constants. It is a Bertotti-Robinson-like
metric. Components of $T_{ab}$ are as follows
\begin{eqnarray*}
T_{0} &=&0 ,
T_{1} =0 , \\
T_{2} &=&a^{2}B^{2} ,  T_{3} =B^{2} .
\end{eqnarray*}
and $R_{ab}$ has the following components
\begin{eqnarray*}
R_{0} &=&B^{2}\cosh ^{2}(A+B\rho) ,
R_{1} =B^{2} , \\
R_{2} &=&0 ,  R_{3} =0 .
\end{eqnarray*}
It has infinite dimensional Lie algebras both for MCs (case (IX))
and RCs (Case (X)) 6 KVs (Case AIIb2(ii)$\alpha _{2}$).

\section{Conclusion}

We have studied the relationship between the Lie symmetries or
collineations of the two second rank tensors, the energy-momentum
and the Ricci tensors, which are mathematically very similar. In
particular, we investigate whether or not this similarity and their
duality in the EFEs is preserved by their collineations also. For
this purpose we have used the framework of cylindrically symmetric
static manifolds. The KVs and RCs of these spaces have been
classified earlier \cite{QZ, KS}. While KVs have a finite
dimensional Lie algebra always, RCs and MCs can admit infinite
dimensional Lie algebra as well. Similarly, RCs and MCs can be
degenerate or non-degenerate. In this way we see that, in all, there
are a total of nine types of relationships between RCs and MCs which
are formulated in a table in Section 3. To show that they are not
just symmetries, to which no solutions of EFEs exist, we have
explicitly constructed examples for all of these cases, except for
the two cases 3.5 and 3.6. For these two cases we have not been able
to provide any example, nor have we managed to prove that they do
not exist. Unless and until the examples for these two cases are
provided the question of ``duality" of the Lie symmetries of the
energy-momentum and the Ricci tensors will remain open.

It is worth while explaining the problem in finding the examples.
Despite the apparent duality of the tensors in the EFEs, there is an
enormous difference in the differential equations defining the
tensors. At the very least, this complicates the equations to the
point that while we can construct the solutions for the cases for
the Ricci tensor, we are unable to do so for the energy-momentum
tensor. It appears to be a distinct possibility that there is no
duality between the two tensors because of the difference in the
differential equations yielding the cases. It may be that the answer
to our question will come by investigating the structure of the two
differential equations. \\ \\
\textbf{Appendix} \\ \\
The tables in the appendix summarize the solutions of Eq.(\ref{RC}).
These are, in fact, obtained by changing the components of the Ricci
tensor in Ref. \cite{KS} by those of the energy-momentum tensor.
Thus the equation numbers in the last columns of these tables refer
to the equations in Ref. \cite{KS}.



\begin{sidewaystable}
\begin{tiny}

\begin{tabular}{ccccc}
&  &  &  &  \\
&  &  &  &  \\
&  &  &  &  \\
&  &  &  &  \\
&  &  &  &  \\
&  &  &  &  \\
&  &  &  &  \\
&  &  &  &  \\ \hline
\multicolumn{5}{|c|}{} \\
\multicolumn{5}{|c|}{\textbf{Tables for the Matter Collineations of
Cylindrically Symmetric Static Spacetimes}} \\
\multicolumn{5}{|c|}{} \\
\hline \multicolumn{4}{|c|}{\textbf{The Non-Degenerate
Energy-Momentum Tensor
}} & \multicolumn{1}{|c|}{\textbf{The Degenerate Energy-Momentum Tensor }} \\
\hline \multicolumn{3}{|c|}{\textbf{Case\thinspace A:\
}$T_{0}^{\prime }\neq 0$} &
\multicolumn{1}{|c}{\textbf{Case\thinspace B:} $T_{0}^{\prime }=0$}
& \multicolumn{1}{|c|}{\textbf{Case\thinspace II: \thinspace
}$T_{1}=0\,$,$\ T_{0}\neq 0$\thinspace ,$\ T_{2}\neq 0\,$,$\
T_{3}\neq 0$} \\ \hline \multicolumn{2}{|c}{\textbf{Case\thinspace
A(I):\ }$\left( \frac{T_{2}}{T_{3} }\right) ^{\prime }\neq 0$} &
\multicolumn{1}{|c}{\textbf{Case\thinspace A(II):\ }$\left(
\frac{T_{2}}{T_{3}}\right) ^{\prime }=0$} & \multicolumn{1}{|c}{} &
\multicolumn{1}{|c|}{} \\ \hline \multicolumn{1}{|c}{\textbf{(a)
}$T_{2}^{\prime }=0\,$, $T_{3}^{\prime }\neq
0$} & \multicolumn{1}{|c}{\textbf{(c) }$T_{2}^{\prime }\neq 0\,$, $%
T_{3}^{\prime }\neq 0$} & \multicolumn{1}{|c}{} &
\multicolumn{1}{|c}{} & \multicolumn{1}{|c|}{} \\ \cline{1-2}
\multicolumn{1}{|c}{\textbf{(b) }$T_{2}^{\prime }\neq 0\,$,
$T_{3}^{\prime }=0$} & \multicolumn{1}{|c}{} & \multicolumn{1}{|c}{}
& \multicolumn{1}{|c}{} & \multicolumn{1}{|c|}{} \\ \hline\hline
\multicolumn{1}{|c}{\textbf{Table 1}} &
\multicolumn{1}{|c}{\textbf{Table 2}} &
\multicolumn{1}{|c}{\textbf{Table 3}} &
\multicolumn{1}{|c}{\textbf{Table 4}} &
\multicolumn{1}{|c|}{\textbf{Table 5}} \\ \hline
\end{tabular}
\end{tiny}
\end{sidewaystable}

\pagebreak

\begin{sidewaystable}
\begin{small}

\begin{tabular}{|lllllll|}
\hline
\multicolumn{7}{|l|}{} \\
\multicolumn{7}{|c|}{\textbf{Table 1: The Non-Degenerate
Case\thinspace
A(I)\quad }$\quad \quad \quad T_{0}^{\prime }\neq 0$} \\
\multicolumn{7}{|c|}{} \\ \hline (I) $\left(
\frac{T_{2}}{T_{3}}\right) ^{\prime }\neq 0$ &
\multicolumn{1}{|l}{(a) $T_{2}^{\prime }=0\,$,$\,$} &
\multicolumn{1}{|l}{ (1) $\alpha =0,\,\,\beta =0$} &
\multicolumn{1}{|l}{(i) $\left( \frac{T_{0}}{T_{3}}\right)
^{^{\prime }}=0$} & \multicolumn{1}{|l}{} & \multicolumn{1}{|l}{} &
\multicolumn{1}{|l|}{7 MCs (Eqs. 16)} \\
& \multicolumn{1}{|l}{$\,\,\,\,\,\,\,\,\,T_{3}^{\prime }\neq 0$} &
\multicolumn{1}{|l}{$\,\,\,\,\,\,\,\,\,$} & \multicolumn{1}{|l}{} &
\multicolumn{1}{|l}{} & \multicolumn{1}{|l}{} & \multicolumn{1}{|l|}{} \\
\hline & \multicolumn{1}{|l}{} & \multicolumn{1}{|l}{} &
\multicolumn{1}{|l}{(ii) $ \left( \frac{T_{0}}{T_{3}}\right)
^{^{\prime }}\neq 0$} & \multicolumn{1}{|l}{} &
\multicolumn{1}{|l}{} & \multicolumn{1}{|l|}{4 MCs  (Eqs. 17)}
\\ \hline & \multicolumn{1}{|l}{} & \multicolumn{1}{|l}{(2) $\alpha
\neq 0,\,\beta =0$} & \multicolumn{1}{|l}{(i) $\alpha >0$} &
\multicolumn{1}{|l}{} & \multicolumn{1}{|l}{} &
\multicolumn{1}{|l|}{3 MCs }
\\ \hline
& \multicolumn{1}{|l}{} & \multicolumn{1}{|l}{} & \multicolumn{1}{|l}{(ii) $%
\alpha <0$} & \multicolumn{1}{|l}{($\alpha $) $\left( \frac{T_{0}}{T_{3}}%
\right) ^{\prime }\neq 0$} & \multicolumn{1}{|l}{} &
\multicolumn{1}{|l|}{3 MCs} \\ \hline & \multicolumn{1}{|l}{} &
\multicolumn{1}{|l}{} & \multicolumn{1}{|l}{} &
\multicolumn{1}{|l}{($\beta $) $\left( \frac{T_{0}}{T_{3}}\right)
^{\prime }=0$} & \multicolumn{1}{|l}{} & \multicolumn{1}{|l|}{4 MCs
(Eqs. 18)} \\ \hline & \multicolumn{1}{|l}{} &
\multicolumn{1}{|l}{(3) $\alpha =0,\,\beta \neq 0$} &
\multicolumn{1}{|l}{} & \multicolumn{1}{|l}{} &
\multicolumn{1}{|l}{} & \multicolumn{1}{|l|}{Similar to (2)} \\
\hline & \multicolumn{1}{|l}{} & \multicolumn{1}{|l}{(4) $\alpha
\neq 0,\,\,\,\,\beta \neq 0$} & \multicolumn{1}{|l}{(i) $\alpha
>0,\,\beta >0$} &
\multicolumn{1}{|l}{($\alpha $) $\beta \int \frac{\sqrt{T_{1}}}{T_{0}}d\rho -%
\frac{T_{3}^{^{\prime }}}{2T_{0}\sqrt{T_{1}}}\neq 0$} & \multicolumn{1}{|l}{(%
$\alpha _{1}$) $\left( \frac{T_{0}}{T_{3}}\right) ^{^{\prime }}=0$}
& \multicolumn{1}{|l|}{4 MCs (Eqs. 19)}
\\ \hline & \multicolumn{1}{|l}{} & \multicolumn{1}{|l}{} &
\multicolumn{1}{|l}{} &
\multicolumn{1}{|l}{} & \multicolumn{1}{|l}{($\alpha _{2}$) $\left( \frac{%
T_{0}}{T_{3}}\right) ^{^{\prime }}\neq 0$} & \multicolumn{1}{|l|}{3 MCs} \\
\hline & \multicolumn{1}{|l}{} &
\multicolumn{1}{|l}{$\,\,\,\,\,\,\,\,$} &
\multicolumn{1}{|l}{} & \multicolumn{1}{|l}{($\beta $) $\beta \int \frac{%
\sqrt{T_{1}}}{T_{0}}d\rho -\frac{T_{3}^{^{\prime
}}}{2T_{0}\sqrt{T_{1}}}=0$} & \multicolumn{1}{|l}{} &
\multicolumn{1}{|l|}{3 MCs} \\ \hline
& \multicolumn{1}{|l}{} & \multicolumn{1}{|l}{} & \multicolumn{1}{|l}{(ii) $%
\alpha >0,\,\beta <0$} & \multicolumn{1}{|l}{} &
\multicolumn{1}{|l}{} & \multicolumn{1}{|l|}{Similar to (i)} \\
\hline
& \multicolumn{1}{|l}{} & \multicolumn{1}{|l}{} & \multicolumn{1}{|l}{(iii) $%
\alpha >0,\,\beta >0$} & \multicolumn{1}{|l}{} &
\multicolumn{1}{|l}{} & \multicolumn{1}{|l|}{Similar to (i)} \\
\hline
& \multicolumn{1}{|l}{} & \multicolumn{1}{|l}{} & \multicolumn{1}{|l}{(iv) $%
\alpha <0,\,\beta <0$} & \multicolumn{1}{|l}{} &
\multicolumn{1}{|l}{} & \multicolumn{1}{|l|}{Similar to (i)} \\
\hline & \multicolumn{1}{|l}{(b) $T_{2}^{\prime }\neq $ $0\,$,} &
\multicolumn{1}{|l}{} & \multicolumn{1}{|l}{} &
\multicolumn{1}{|l}{} &
\multicolumn{1}{|l}{} & \multicolumn{1}{|l|}{Similar to (a)} \\
& \multicolumn{1}{|l}{$\,\,\,\,\,\,\,\,\,\,\,T_{3}^{\prime }=0$} &
\multicolumn{1}{|l}{} & \multicolumn{1}{|l}{} &
\multicolumn{1}{|l}{} & \multicolumn{1}{|l}{} &
\multicolumn{1}{|l|}{} \\ \hline\hline
Definitions &  & $\alpha =\frac{T_{0}}{\sqrt{T_{1}}}\left( \frac{%
T_{0}^{\prime }}{2T_{0}\sqrt{T_{1}}}\right) ^{\prime }$ & $k_{1}=-\frac{%
T_{0}^{\prime }}{2T_{0}\sqrt{T_{1}}}$ &  &  &  \\
&  & $\beta =\frac{T_{3}}{\sqrt{T_{1}}}\left( \frac{T_{3}^{\prime }}{2T_{3}%
\sqrt{T_{1}}}\right) ^{\prime }$ & $k_{2}=-\frac{T_{3}^{\prime }}{2T_{3}%
\sqrt{T_{1}}}$ &  &  &  \\ \hline
\end{tabular}
\end{small}
\end{sidewaystable}

\pagebreak
\begin{sidewaystable}
\begin{small}

\begin{tabular}{|l|ll|ll|ll|l|l|}
\hline
\multicolumn{9}{|l|}{} \\
\multicolumn{9}{|c|}{\textbf{Table 2:\ The Non-Degenerate
Case\thinspace A(I)\quad }$\quad \quad \quad T_{0}^{\prime }\neq 0$
\thinspace \thinspace \thinspace \thinspace \thinspace \thinspace
\thinspace \thinspace \thinspace
\thinspace \thinspace (continued)} \\
\multicolumn{9}{|c|}{} \\ \hline
(I) $\left( \frac{T_{2}}{T_{3}}\right) ^{\prime }\neq 0$ & (c) & $%
T_{2}^{\prime }\neq $ $0\,$, & (1) & $\left( \frac{T_{2}^{\prime
}}{2T_{2} \sqrt{T_{1}}}\right) ^{\prime }\neq 0$ , & (i) & $\left(
\frac{T_{2}}{T_{0}}
\right) ^{^{\prime }}=0,$ &  & 4 MCs (Eqs. 16) \\
&  & $T_{3}^{\prime }\neq 0$ &  & $\left( \frac{T_{3}^{\prime
}}{2T_{3}\sqrt{ T_{1}}}\right) ^{\prime }\neq 0$ &  & $\left(
\frac{T_{3}}{T_{0}}\right) ^{^{\prime }}\neq 0$ &  &  \\ \hline &  &
&  &  & (ii) & $\left( \frac{T_{2}}{T_{0}}\right) ^{^{\prime }}\neq
0,$
&  & Similar to (i) \\
&  &  &  &  &  & $\left( \frac{T_{3}}{T_{0}}\right) ^{^{\prime }}=0$
&  &
\\ \hline
&  &  &  &  & (iii) & $\left( \frac{T_{2}}{T_{0}}\right) ^{^{\prime
}}\neq
0, $ &  & 3 MCs \\
&  &  &  &  &  & $\left( \frac{T_{3}}{T_{0}}\right) ^{^{\prime
}}\neq 0$ & &  \\ \hline &  &  & (2) & $\left( \frac{T_{2}^{\prime
}}{2T_{2}\sqrt{T_{1}}}\right) ^{\prime }=0$ , & (i) & $\left(
\frac{T_{2}}{T_{0}}\right) ^{^{\prime }}=0,$
&  & 5 MCs (Eqs. 22) \\
&  &  &  & $\left( \frac{T_{3}^{\prime }}{2T_{3}\sqrt{T_{1}}}\right)
^{\prime }=0$ &  & $\left( \frac{T_{3}}{T_{0}}\right) ^{^{\prime
}}\neq 0$ & &  \\ \hline &  &  &  &  & (ii) & $\left(
\frac{T_{2}}{T_{0}}\right) ^{^{\prime }}\neq 0,$
&  & Similar to (i) \\
&  &  &  &  &  & $\left( \frac{T_{3}}{T_{0}}\right) ^{^{\prime }}=0$
&  &
\\ \hline
&  &  &  &  & (iii) & $\left( \frac{T_{2}}{T_{0}}\right) ^{^{\prime
}}\neq 0, $ & ($\alpha $) $\left( \frac{T_{0}^{\prime
}}{2T_{0}\sqrt{T_{1}}}\right)
^{\prime }=0$ & 4 MCs (Eqs. 23) \\
&  &  &  &  &  & $\left( \frac{T_{3}}{T_{0}}\right) ^{^{\prime
}}\neq 0$ & &  \\ \hline
&  &  &  &  &  &  & ($\beta $) $\left( \frac{T_{0}^{\prime }}{2T_{0}\sqrt{%
T_{1}}}\right) ^{\prime }\neq 0$ & 3 MCs \\ \hline
\end{tabular}
\end{small}
\end{sidewaystable}

\pagebreak
\begin{sidewaystable}
\begin{tiny}

\begin{tabular}{|lllllll|}
\hline
\multicolumn{7}{|l|}{} \\
\multicolumn{7}{|c|}{\textbf{Table 3:\ The Non-Degenerate
Case\thinspace
A(II)\quad }$\quad \quad \quad T_{0}^{\prime }\neq 0$} \\
\multicolumn{7}{|c|}{} \\ \hline (II) $\left(
\frac{T_{2}}{T_{3}}\right) ^{\prime }=0$ & \multicolumn{1}{|l}{ (a)
$\left( \frac{T_{2}^{\prime }}{2T_{2}\sqrt{T_{1}}}\right) ^{\prime
}\neq 0$} & \multicolumn{1}{|l}{(1) $\left(
\sqrt{\frac{T_{0}}{T_{2}}}\right) ^{\prime }=0$} &
\multicolumn{1}{|l}{} & \multicolumn{1}{|l}{} &
\multicolumn{1}{|l}{} & \multicolumn{1}{|l|}{6 MCs (Eqs. 24)} \\
\hline
& \multicolumn{1}{|l}{} & \multicolumn{1}{|l}{(2) $\left( \sqrt{\frac{T_{0}}{%
T_{2}}}\right) ^{\prime }\neq 0$} & \multicolumn{1}{|l}{} &
\multicolumn{1}{|l}{} & \multicolumn{1}{|l}{} &
\multicolumn{1}{|l|}{4 MCs (Eqs. 25)} \\
\hline
& \multicolumn{1}{|l}{(b) $\left( \frac{T_{2}^{\prime }}{2T_{2}\sqrt{T_{1}}}%
\right) ^{\prime }=0$} & \multicolumn{1}{|l}{(1) $\alpha \neq 0$} &
\multicolumn{1}{|l}{(i) $\left( \frac{T_{2}}{T_{0}}\right)
^{^{\prime }}\neq
0$} & \multicolumn{1}{|l}{($\alpha $) $\left( \frac{T_{0}^{\prime }}{2T_{0}%
\sqrt{T_{1}}}\right) ^{\prime }\neq 0$} & \multicolumn{1}{|l}{} &
\multicolumn{1}{|l|}{4 MCs (Eqs. 26)}
\\ \hline & \multicolumn{1}{|l}{} & \multicolumn{1}{|l}{} &
\multicolumn{1}{|l}{} &
\multicolumn{1}{|l}{($\beta $) $\left( \frac{T_{0}^{\prime }}{2T_{0}\sqrt{%
T_{1}}}\right) ^{\prime }=0$} & \multicolumn{1}{|l}{} &
\multicolumn{1}{|l|}{5 MCs (Eqs. 27)}
\\ \hline
& \multicolumn{1}{|l}{} & \multicolumn{1}{|l}{} & \multicolumn{1}{|l}{(ii) $%
\left( \frac{T_{2}}{T_{0}}\right) ^{^{\prime }}=0$} &
\multicolumn{1}{|l}{} & \multicolumn{1}{|l}{} &
\multicolumn{1}{|l|}{10 MCs (Eqs. 28)}
\\ \hline & \multicolumn{1}{|l}{} & \multicolumn{1}{|l}{(2) $\alpha
=0$ ,} &
\multicolumn{1}{|l}{(i) $\left( \frac{\left( \sqrt{T_{0}}\right) ^{\prime }}{%
\sqrt{T_{1}}}\right) ^{\prime }=0$\thinspace} &
\multicolumn{1}{|l}{} & \multicolumn{1}{|l}{} &
\multicolumn{1}{|l|}{10 MCs (Eqs. 29)}
\\ \hline
& \multicolumn{1}{|l}{} & \multicolumn{1}{|l}{} & \multicolumn{1}{|l}{(ii) $%
\left( \frac{\left( \sqrt{T_{0}}\right) ^{\prime
}}{\sqrt{T_{1}}}\right)
^{\prime }\neq 0$\thinspace} & \multicolumn{1}{|l}{($\alpha $) $\left[ \frac{%
T_{0}}{2\sqrt{T_{1}}}\left( \frac{T_{0}^{\prime
}}{T_{0}\sqrt{T_{1}}}\right) ^{\prime }\right] ^{\prime }=0$} &
\multicolumn{1}{|l}{($\alpha _{1}$) $\eta
=0$} & \multicolumn{1}{|l|}{6 MCs (Eqs. 30)} \\
\hline & \multicolumn{1}{|l}{} & \multicolumn{1}{|l}{} &
\multicolumn{1}{|l}{} & \multicolumn{1}{|l}{} &
\multicolumn{1}{|l}{($\alpha _{2}$) $\eta \neq 0$} &
\multicolumn{1}{|l|}{6 MCs (Eqs. 31)}
\\ \hline & \multicolumn{1}{|l}{} & \multicolumn{1}{|l}{} &
\multicolumn{1}{|l}{} & \multicolumn{1}{|l}{($\beta $) $\left[
\frac{T_{0}}{2\sqrt{T_{1}}}\left( \frac{T_{0}^{\prime
}}{T_{0}\sqrt{T_{1}}}\right) ^{\prime }\right] ^{\prime }\neq 0$} &
\multicolumn{1}{|l}{} & \multicolumn{1}{|l|}{4 MCs (Eqs. 32)} \\
\hline\hline Definitions &  & $\alpha =\frac{T_{2}^{\prime
}}{T_{2}\sqrt{T_{1}}}$ &  &  &
$\eta =\frac{T_{0}}{2\sqrt{T_{1}}}\left( \frac{T_{0}^{\prime }}{T_{0}\sqrt{%
T_{1}}}\right) ^{\prime }$ &  \\
&  &  &  &  &  &  \\ \hline
\end{tabular}
\end{tiny}
\end{sidewaystable}

\pagebreak

\begin{sidewaystable}
\begin{tiny}

\begin{tabular}{|lllllll|}
\hline
\multicolumn{7}{|l|}{} \\
\multicolumn{7}{|c|}{\textbf{Table 4:\ The Non-Degenerate
Case\thinspace
B\quad }$\quad \quad \quad T_{0}^{\prime }=0$} \\
\multicolumn{7}{|c|}{} \\ \hline (I) $T_{2}^{\prime }=$ $0\,$,
$T_{3}^{\prime }=0$ & \multicolumn{1}{|l}{} & \multicolumn{1}{|l}{}
& \multicolumn{1}{|l}{} & \multicolumn{1}{|l}{} &
\multicolumn{1}{|l}{} & \multicolumn{1}{|l|}{10 MCs (Eqs. 33)} \\
\hline (II) $T_{2}^{\prime }=$ $0\,$, $T_{3}^{\prime }\neq 0$ &
\multicolumn{1}{|l}{
(a) $\left[ \frac{T_{3}}{\sqrt{T_{1}}}\left( \frac{T_{3}^{\prime }}{2T_{3}%
\sqrt{T_{1}}}\right) ^{\prime }\right] ^{\prime }\neq 0$} &
\multicolumn{1}{|l}{} & \multicolumn{1}{|l}{} &
\multicolumn{1}{|l}{} & \multicolumn{1}{|l}{} &
\multicolumn{1}{|l|}{4 MCs (Eqs. 34)}
\\ \hline
& \multicolumn{1}{|l}{(b) $\left[ \frac{T_{3}}{\sqrt{T_{1}}}\left( \frac{%
T_{3}^{\prime }}{2T_{3}\sqrt{T_{1}}}\right) ^{\prime }\right]
^{\prime }=0$} & \multicolumn{1}{|l}{(1) $k_{1}>0$} &
\multicolumn{1}{|l}{} & \multicolumn{1}{|l}{} &
\multicolumn{1}{|l}{} & \multicolumn{1}{|l|}{6 MCs (Eqs. 35)} \\
\hline & \multicolumn{1}{|l}{} & \multicolumn{1}{|l}{(2) $k_{1}<0$}
& \multicolumn{1}{|l}{} & \multicolumn{1}{|l}{} &
\multicolumn{1}{|l}{} & \multicolumn{1}{|l|}{Similar to (1)} \\
\hline & \multicolumn{1}{|l}{} & \multicolumn{1}{|l}{(3) $k_{1}=0$}
& \multicolumn{1}{|l}{} & \multicolumn{1}{|l}{} &
\multicolumn{1}{|l}{} & \multicolumn{1}{|l|}{6 MCs (Eqs. 36)} \\
\hline (III) $T_{2}^{\prime }\neq $ $0\,$, $T_{3}^{\prime }=0$ &
\multicolumn{1}{|l}{} & \multicolumn{1}{|l}{} &
\multicolumn{1}{|l}{} & \multicolumn{1}{|l}{} &
\multicolumn{1}{|l}{} & \multicolumn{1}{|l|}{Similar to (II)} \\
\hline (IV) $T_{2}^{\prime }\neq $ $0\,$, $T_{3}^{\prime }\neq 0$ &
\multicolumn{1}{|l}{(a) $\left[ \frac{T_{2}}{\sqrt{T_{1}}}\left(
\frac{ T_{2}^{\prime }}{2T_{2}\sqrt{T_{1}}}\right) ^{\prime }\right]
^{\prime }\neq 0$} & \multicolumn{1}{|l}{(1) $\left(
\frac{T_{2}}{T_{3}}\right) ^{^{\prime }}\neq 0$} &
\multicolumn{1}{|l}{} & \multicolumn{1}{|l}{} &
\multicolumn{1}{|l}{} & \multicolumn{1}{|l|}{3 MCs} \\ \hline &
\multicolumn{1}{|l}{} & \multicolumn{1}{|l}{(2) $\left(
\frac{T_{2}}{T_{3}} \right) ^{^{\prime }}=0$} &
\multicolumn{1}{|l}{} & \multicolumn{1}{|l}{} &
\multicolumn{1}{|l}{} & \multicolumn{1}{|l|}{4 MCs (Eqs. 37)}
\\ \hline
& \multicolumn{1}{|l}{(b) $\left[ \frac{T_{2}}{\sqrt{T_{1}}}\left(
\frac{ T_{2}^{\prime }}{2T_{2}\sqrt{T_{1}}}\right) ^{\prime }\right]
^{\prime }=0$} & \multicolumn{1}{|l}{(1) $k_{3}>0$} &
\multicolumn{1}{|l}{(i) $\left[ \frac{ T_{3}}{\sqrt{T_{1}}}\left(
\frac{T_{3}^{\prime }}{2T_{3}\sqrt{T_{1}}}\right) ^{\prime }\right]
^{\prime }\neq 0$} & \multicolumn{1}{|l}{} & \multicolumn{1}{|l}{} &
\multicolumn{1}{|l|}{4 MCs (Eqs. 38)}
\\ \hline
& \multicolumn{1}{|l}{} & \multicolumn{1}{|l}{} &
\multicolumn{1}{|l}{(ii) $ \left[ \frac{T_{3}}{\sqrt{T_{1}}}\left(
\frac{T_{3}^{\prime }}{2T_{3}\sqrt{T_{1}}}\right) ^{\prime }\right]
^{\prime }=0$} & \multicolumn{1}{|l}{} & \multicolumn{1}{|l}{} &
\multicolumn{1}{|l|}{4 MCs (Eqs. 38)}
\\ \hline & \multicolumn{1}{|l}{} & \multicolumn{1}{|l}{(2)
$k_{3}<0$} & \multicolumn{1}{|l}{} & \multicolumn{1}{|l}{} &
\multicolumn{1}{|l}{} & \multicolumn{1}{|l|}{Similar to (1)} \\
\hline & \multicolumn{1}{|l}{} & \multicolumn{1}{|l}{(3) $k_{3}=0$}
&
\multicolumn{1}{|l}{(i) $\left[ \frac{T_{3}}{\sqrt{T_{1}}}\left( \frac{%
T_{3}^{\prime }}{2T_{3}\sqrt{T_{1}}}\right) ^{\prime }\right]
^{\prime }\neq 0$} & \multicolumn{1}{|l}{} & \multicolumn{1}{|l}{} &
\multicolumn{1}{|l|}{4 MCs (Eqs. 38)}
\\ \hline
& \multicolumn{1}{|l}{} & \multicolumn{1}{|l}{} & \multicolumn{1}{|l}{(ii) $%
\left[ \frac{T_{3}}{\sqrt{T_{1}}}\left( \frac{T_{3}^{\prime }}{2T_{3}\sqrt{%
T_{1}}}\right) ^{\prime }\right] ^{\prime }=0$} & \multicolumn{1}{|l}{($%
\alpha $) $k_{1}>0$} & \multicolumn{1}{|l}{} &
\multicolumn{1}{|l|}{4 MCs (Eqs. 38)} \\ \hline &
\multicolumn{1}{|l}{} & \multicolumn{1}{|l}{} &
\multicolumn{1}{|l}{} & \multicolumn{1}{|l}{($\beta $) $k_{1}<0$} &
\multicolumn{1}{|l}{} & \multicolumn{1}{|l|}{Similar to (2)} \\
\hline & \multicolumn{1}{|l}{} & \multicolumn{1}{|l}{} &
\multicolumn{1}{|l}{} & \multicolumn{1}{|l}{($\gamma $) $k_{1}=0$} &
\multicolumn{1}{|l}{($\gamma _{1}$) $\left(
\frac{T_{3}}{T_{2}}\right) ^{^{\prime }}\neq 0$} &
\multicolumn{1}{|l|}{4 MCs (Eqs. 39)}
\\ \hline & \multicolumn{1}{|l}{} & \multicolumn{1}{|l}{} &
\multicolumn{1}{|l}{} &
\multicolumn{1}{|l}{} & \multicolumn{1}{|l}{($\gamma _{2}$) $\left( \frac{%
T_{3}}{T_{2}}\right) ^{^{\prime }}=0$} & \multicolumn{1}{|l|}{7 MCs
(Eqs. 40)} \\ \hline\hline Definitions &  &
$k_{1}=\frac{T_{3}}{\sqrt{T_{1}}}\left( \frac{T_{3}^{\prime
}}{2T_{3}\sqrt{T_{1}}}\right) ^{\prime }$ &  &  &  &  \\
&  & $k_{3}=\frac{T_{2}}{\sqrt{T_{1}}}\left( \frac{T_{2}^{\prime }}{2T_{2}%
\sqrt{T_{1}}}\right) ^{\prime }$ &  &  &  &  \\ \hline
\end{tabular}
\end{tiny}

\end{sidewaystable}

\pagebreak

\begin{sidewaystable}
\begin{small}

\begin{tabular}{|l|l|l|l|l|}
\hline
\multicolumn{5}{|l|}{} \\
\multicolumn{5}{|c|}{\textbf{Table 5:\ The Degenerate Case\thinspace
II \thinspace \thinspace \thinspace \thinspace \thinspace \thinspace
\thinspace \thinspace \thinspace \thinspace \thinspace \thinspace
\thinspace \thinspace \thinspace \thinspace \thinspace
}$T_{1}=0\,$,$\ T_{0}\neq 0$\thinspace ,$\
T_{2}\neq 0\,$,$\ T_{3}\neq 0$} \\
\multicolumn{5}{|l|}{} \\ \hline (A) $T_{0}^{\prime}=0$ &
$T_{2}^{\prime}\neq 0\,,\,T_{3}^{\prime}\neq 0$ & (1) $\left(
\frac{T_{2}^{\prime
}T_{3}}{T_{2}T_{3}^{\prime }}\right) ^{\prime }=0$ &  & 4 MCs (Eqs. 41) \\
\hline &  & (2) $\left( \frac{T_{2}^{\prime
}T_{3}}{T_{2}T_{3}^{\prime }}\right)^{\prime }\neq 0$
&  & 3 MCs \\ \hline & Otherwise &  &  & Infinitely many MCs \\
\hline (B) $T_{0}^{\prime}\neq 0$ & (a) $\left(
\frac{T_{0}}{T_{2}}\right) ^{\prime }=0,\left(
\frac{T_{0}}{T_{3}}\right) ^{\prime }=0$ &  &  & 10 MCs (Eqs. 42) \\
\hline & (b) $\left( \frac{T_{0}}{T_{2}}\right) ^{\prime }=0,\left(
\frac{T_{0}}{T_{3}}\right) ^{\prime }\neq 0$ & (1) $\left(
\frac{T_{3}^{\prime }T_{0}}{T_{3}T_{0}^{\prime }}\right) ^{\prime
}=0$ &  & 5 MCs (Eqs. 43) \\ \hline& & (2) $\left(
\frac{T_{3}^{\prime }T_{0}}{T_{3}T_{0}^{\prime }}\right)
^{\prime }\neq 0$ &  & 4 MCs (Eqs. 44) \\
\hline & (c) $\left( \frac{T_{0}}{T_{2}}\right) ^{\prime }\neq
0,\left( \frac{T_{0}}{T_{3}}\right) ^{\prime }=0$ &  &  & Similar to
(b) \\ \hline & (d) $\left( \frac{T_{0}}{T_{2}}\right) ^{\prime
}\neq 0,\left( \frac{T_{0}}{T_{3}}\right) ^{\prime }\neq 0$ & (1)
$\left( \frac{T_{0}^{\prime }T_{2}}{T_{0}T_{2}^{\prime }}\right)
^{\prime }\neq 0, \left( \frac{T_{0}^{\prime
}T_{3}}{T_{0}T_{3}^{\prime }}\right) ^{\prime }\neq 0$ & (i) $
\left( \frac{T_{2}}{T_{3}}\right) ^{\prime }=0$ & 4 MCs (Eqs. 45) \\
\hline
&  &  & (ii) $\left( \frac{T_{2}}{T_{3}}\right) ^{\prime }\neq 0$ & 3 MCs \\
\hline &  & (2) $\left( \frac{T_{0}^{\prime
}T_{2}}{T_{0}T_{2}^{\prime }}\right) ^{\prime }\neq 0\,$,$\,\left(
\frac{T_{0}^{\prime }T_{3}}{T_{0}T_{3}^{\prime }}\right) ^{\prime
}=0$ &  & 4 MCs (Eqs. 45)
\\ \hline
&  & (3) $\left( \frac{T_{0}^{\prime }T_{2}}{T_{0}T_{2}^{\prime
}}\right) ^{\prime }=0\,$,$\,\left( \frac{T_{0}^{\prime
}T_{3}}{T_{0}T_{3}^{\prime }} \right) ^{\prime }\neq 0$ &  & Similar
to (2) \\ \hline &  & (4) $\left( \frac{T_{0}^{\prime
}T_{2}}{T_{0}T_{2}^{\prime }}\right) ^{\prime }=0\,$,$\,\left(
\frac{T_{0}^{\prime }T_{3}}{T_{0}T_{3}^{\prime }} \right)
^{\prime}=0$ & (i) $\left( \frac{T_{2}}{T_{3}}\right)
^{\prime }=0$ & 5 MCs (Eqs. 46) \\
\hline
&  &  & (ii) $\left( \frac{T_{2}}{T_{3}}\right) ^{\prime }\neq 0$ & 4 MCs \\
\hline
\end{tabular}
\end{small}

\end{sidewaystable}




\acknowledgments


\bigskip

KS acknowledges a research grant from the Higher Education
Commission of Pakistan. MZ gratefully acknowledges the Internal
Research Grant: IG/SCI/DOMS/10/01 by Sultan Qaboos University,
Sultanate of Oman.


\end{document}